 \documentstyle[12pt]{article}
 \def\be{\begin{eqnarray}}
 \def\ee{\end{eqnarray}}
 \def\gbe{{\gamma_m^{(0)}/\beta_0}}
 \def\gzero{\gamma_m^{(0)}}
 \def\lll{{\cal L}}
 
\begin{document}
 \thispagestyle{empty}
 \begin{flushright}
 {Alberta Thy-38-93}    \\[2mm]
 {September 1993}           \\
 \end{flushright}
 \vspace{1cm}
 \begin{center}
 {\bf \large
 Two-loop renormalization group analysis of hadronic decays \\[3mm]
 of a charged Higgs boson }
 \end{center}
 \vspace{1cm}
 \begin{center}
 {Andrzej Czarnecki\footnote{Address after September 30th, 1993:
 Institut f\"ur Physik, Johannes Gutenberg-Universit\"at,
 \mbox{D-55099} Mainz, Germany}}\\
 \vspace{.3cm}
 {\em Department of Physics, University of Alberta, \\
      Edmonton, Canada T6G 2J1}
 \end{center}
 \begin{center}
 and
 \end{center}
 \begin{center}
 {Andrei I.~Davydychev\footnote{On leave from
 Institute for Nuclear Physics,
 Moscow State University, 119899, Moscow, Russia.
 E-mail address: davyd@vsfys1.fi.uib.no}}\\
 \vspace{.3cm}
 {\em Department of Physics, University of Bergen,\\
      All\'{e}gaten 55, N-5007 Bergen, Norway}
 \end{center}
 \hspace{3in}
 \begin{abstract}
    We calculate next-to-leading QCD corrections to the decay $H^+
 \rightarrow u\bar d$ for generic up and down quarks in the final
state.
 A recently developed algorithm for evaluation of massive two-loop
 Feynman diagrams is employed to calculate renormalization constants
of
 the charged Higgs boson.  The origin and summation of large
logarithmic
 corrections to the decay rate of the top quark into a lighter
charged
 Higgs boson is also explained.
 \end{abstract}

 \vspace{15mm}

 \begin{center} PACS numbers:  14.80.Gt, 12.38.Cy, 12.15.Cc
 \end{center}

 \newpage
\renewcommand{\topfraction}{0.9}
\renewcommand{\bottomfraction}{0.}
\setcounter{topnumber}{10}
\setcounter{bottomnumber}{0}
\setcounter{totalnumber}{10}
 \setcounter{page}{1}

 \section{Introduction}
 The accelerators planned to be
    built in  the near  future will  provide an  insight into
physics at TeV
 energy scale and thus probe  a region especially interesting from
the point
 of  view  of  electroweak  interactions.   Therefore  we observe
recently an
 increased interest in  various aspects of  phenomenology of the
electroweak
 symmetry  breaking,  in  the  framework  of  the  Standard  Model
and   its
 extensions, which predict one or more  doublets of Higgs bosons.

 As  far as
    the experimental detection of the Higgs particles is concerned,
detailed
 knowledge of their decay properties  is of special interest. In  the
present
 paper we concentrate on hadronic  decays of charged Higgs bosons,
predicted
 e.g.  by  the  Minimal  Supersymmetric  Model  (see
ref.~\cite{hunter} for a
 review and  further references).  In the  case of  the Standard
Model Higgs
 boson hadronic decays have been analyzed  in a number of
publications.   QCD
  and QED  corrections were  first calculated  in \cite{bl80},  where
it  was
 noted that as the ratio of mass  of the decaying scalar particle to
that  of
  fermions in  the final  state increases,  the one-loop  corrections
diverge
 logarithmically. This problem was solved by renormalizing mass of
quarks  at
 the  energy  scale  equal  to  the  mass  of  the decaying particle
and thus
 absorbing the large correction into  the tree level decay rate
expressed in
 terms of the running mass. This approach was subsequently extended
by  means
 of the renormalization group technique in ref.~\cite{inami,sakai}
(see  also
 ref.~\cite{hunter}  for   a  clear   summary),  where
next-to-leading  QCD
  corrections were summed up  in the case of  large mass of the
Higgs boson.
 The  effect  of   three-loop  QCD  corrections   was  first
calculated   in
 ref.~\cite{gorish}   and   further   analyzed   in
ref.~\cite{gorish90-91}.
 Leading  logarithmic  approximation  and  various  ways  to
parametrize the
 next-to-leading   corrections   to   the   Standard   Model   Higgs
boson
 are  subject   of  several   recent  studies
\cite{drees90,ad91,kataev92}.

 In the minimal extension of the Standard Model suggested by
 supersymmetry there are two Higgs boson doublets, and one of the
charged
 modes becomes physical. It is therefore of great importance to look
at
 the phenomenology of such charged scalar particles, since their
discovery
 would
 give information about the theory underlying the Standard Model.
 Various radiative corrections to hadronic decays of charged scalar
particles
 have been calculated and published recently. One-loop QCD
corrections and
 the  sum of leading logarithms can be found
 in \cite{mend90,mend91} and the leading electroweak effects in the
limit
 of large mass of the top quark in \cite{denner90,yang93} (see also
 \cite{diaz93}).

 The purpose of the present paper is to apply the renormalization
group
 technique to calculate next-to-leading QCD corrections to the
process
 $H^+ \rightarrow u \bar d$, where $u$ and $d$ represent generic up
and
 down type quarks respectively. In section \ref{formalism} we
describe
 the theoretical framework of this calculation, and in \ref{rencon}
we
 present the calculation of the two-loop mass and wave function
 renormalization constants of a charged Higgs boson. The results for
 the corrected rate of the Higgs decay are presented in section
 \ref{result}. In section \ref{top} we digress to discuss an
application
 of the summing of leading logarithms to the closely related process
 $t\rightarrow H^+b$. Our conclusions are given in section
\ref{summ}.

 \section{Decay rate and operator product expansion}
 \label{formalism}

 As has been demonstrated in refs.~\cite{inami,sakai},
 the hadronic decay rate of a Higgs boson can be represented by
 \footnote{We shall omit ``plus'' in notations
 related to $H^+$; for example, $m_H \equiv m_{H^+}$,
 $\Gamma_H \equiv \Gamma_{H^+}$, etc.}
 \be
 \Gamma\left(H^+\rightarrow u\bar d \; \right)
 \equiv \Gamma_H
 = {g^2\over m_H} \;
 \mbox{Im}\; C_0(q^2),
 \ee
 where $C_0$ is the coefficient of the unit operator in the operator
 product expansion of the correlator function of the scalar
 currents
 \footnote{The sum on the r.h.s. of (2) goes over the indices $d$ and
$l$,
 where $d$ denotes the canonical dimension of a local operator
 ${\cal O}^l_d$ while $l$ labels independent operators with the same
$d$;
 there is only one (unit) operator with $d=0$: ${\cal O}_0 = 1$.},
 \be
 \mbox{i} \int \mbox{d}^D x \; e^{{\rm i}\;(qx)} \;   %\mbox\small
      \mbox{T}\left[ J_H(x)J_H(0)\right]=\sum_{d,l}
      C^l_d(q^2){\cal O}^l_d,
 \ee
 and the scalar current in the present case is defined by
 \be
 J_H = Z_1^\prime \; {1\over 2m_W} \;
 \bar u \left(a R+  b L \right) d.
 \ee
 In this formula $Z_1^\prime$ denotes the renormalization constant of
 the charged Higgs--fermion vertex,
  $R$ and $L$ are the chiral projection operators,
 $R=(1+\gamma_5)/2$ and $ L=(1-\gamma_5)/2$,
 and the coefficients $a$ and $b$
 depend on the specific model. We will consider two models
characterized
 by the absence of flavour changing neutral currents, described in
detail
 in \cite{hunter}, where references to original papers can also be
found.
 In model I we have
 \be
 a = \sqrt{2} \; m_u\cot\beta, \qquad b = -\sqrt{2} \; m_d\cot\beta,
 \label{eq:abI}
 \ee
 whereas in model II, which corresponds to the Higgs sector of
 the Minimal Supersymmetric Standard Model:
 \be
 a = \sqrt{2} \; m_u\cot\beta, \qquad b = \sqrt{2} \; m_d\tan\beta,
 \label{eq:abII}
 \ee
 and $\tan\beta$ is the ratio of vacuum expectation values of the two
 Higgs doublets.

 We now describe the procedure of calculating next-to-leading
 QCD contributions to the coefficient
 function $C_0(q^2)$.
 Following ref.~\cite{inami} and using methods described in
\cite{gross}
 one derives the renormalization group equation for the
  space-like values of the argument of $C_0$ ($q^2<0$).
 The solution to this equation is found in form of an expansion in
 the QCD coupling constant $g_s$ and in the ratios of masses
 $m_{u,d}/m_H$ (below we shall also use a common notation for the
quark
 masses, $m_q \equiv m_{u,d}$). Keeping only first two terms
 in the mass expansion is
 justified in the region far above the threshold of the $u \bar{d}$
 production; near the threshold not only this expansion is
insufficient,
 but also the perturbative treatment of the vacuum polarization
diagrams
 is questionable.

 While the general method follows closely ref.~\cite{inami} and needs
 no further discussion here, we want to concentrate on the novel
 feature of the present calculation, namely on the computation of the
 renormalization constants of the charged Higgs boson.

 \section{Derivation of renormalization constants}
 \label{rencon}

\begin{figure}[h]
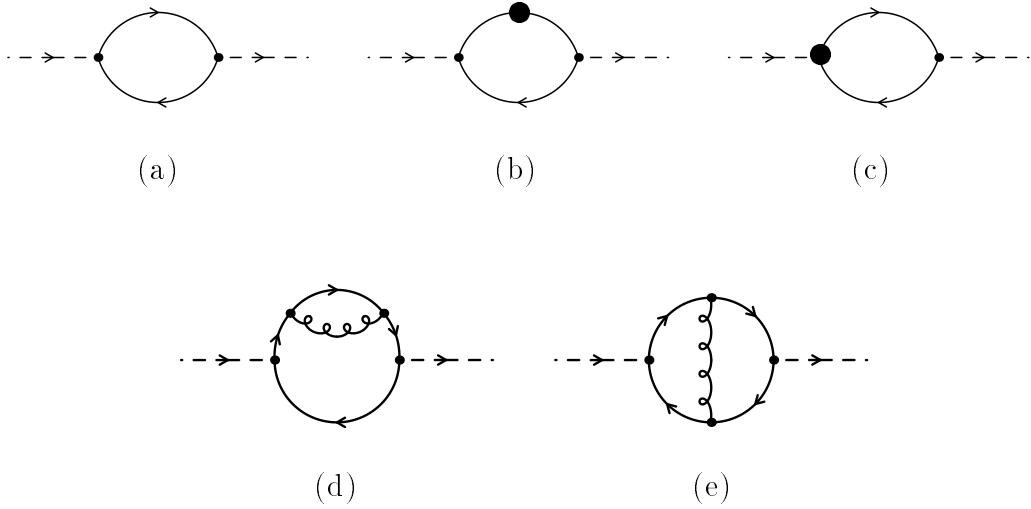

\caption{Types of diagrams contributing to the wave function and mass
renormalization of the charged Higgs boson: one-loop diagram (a),
quark propagator  (b) and  vertex  (c) counterterms,
and reducible (d) and irreducible (e) two-loop diagrams.}
\label{fig:diagr}
\end{figure}

 We work in $D=4-\omega $ dimensional space, considering $\gamma_5$
to be
 anticommuting with other $\gamma$-matrices, $\gamma_5^2 = 1$.
 We may use such a scheme because there is no anomaly
 problem in the case considered (see, e.g., ref.~\cite{gamma_5}).
 Solution of the
 renormalization group equation requires knowledge of $1/\omega$
 poles of the quantities
 \be
 S = g^{-2} (1-Z_3^\prime) \hspace{1cm} \mbox{and} \hspace{1cm}
 T = g^{-2} (1-Z_{m_{{}_{H}}} Z_3^\prime) ,
 \ee
 with $Z_3^\prime$ and $Z_{m_H}$ denoting
 wave function and mass renormalization constants of the charged
Higgs field.
 Hence we have to calculate divergent parts of the self energy
diagrams
 as depicted in figure~\ref{fig:diagr}.
 It has to be noted that in the present case of two
 different masses of quarks in the loop, diagrams corresponding to
figs.
 \ref{fig:diagr}(b) and \ref{fig:diagr}(d) should be considered
together
 with their counterparts with
 corrections on the other quark line. Their sum is
 of course symmetric under $m_u\leftrightarrow m_d$.

 A modification of the method developed in ref.~\cite{ad(s)t93}
enabled
 us to obtain exact expressions for divergent parts of all diagrams
 presented in figure~\ref{fig:diagr}, that are valid for any values
of the
 external momentum $k$, $m_u$ and $m_d$.
 Since in the present paper we are mainly interested in the expansion
 in $m_q^2/m_H^2$, we expand the
 relevant integrals in $m_q^2/k^2$, keeping the $k^2$ and the
 $k^2 (m_q^2/k^2) = m_q^2$ terms only. In the one-loop order
 (see figure~\ref{fig:diagr}(a)) we get
 \be
  -{\mbox{i} g^2 N_{\! C}\over (4\pi)^2 2m_W^2\omega}
 \left\{ 4abm_um_d+(a^2+b^2)\left(2(m_u^2+m_d^2)-k^2\right) \right\},
 \ee
 while the sum of all two-loop-order contributions
 (see figures~\ref{fig:diagr}(b,c,d,e) together with counterparts)
yields
 \be
 \label{2loop-sum}
 {\mbox{i} g^2 g_s^2 N_{\! C}C_F\over (4\pi)^4 4m_W^2\omega}\left\{
 (a^2+b^2)\left[ {12\over\omega}\left(4(m_u^2+m_d^2)-k^2\right)+
 \left( 5k^2-8(m_u^2+m_d^2)\right)\right]
 \right. \nonumber\\ \left.
 +16abm_um_d\left({6\over\omega}-1\right)
 \right\}.
 \ee
 In the above formulae the colour factors are $N_{\! C}=3$ and
$C_F=4/3$.
 It is remarkable that the terms containing $\ln(m_q^2)$ and
$\ln(-k^2)$
 (occurring in the expressions for separate diagrams) disappear in
the
 whole sum (\ref{2loop-sum}). In particular, this fact enables us to
 consider analytic continuation to time-like values of the momentum
 without difficulty.
 It should be also noted that the factors of
 $ \pi^{-\omega/2} \Gamma(1+\frac{1}{2} \omega) $
 are included into the definition of coupling constants $g^2$ and
$g_s^2$,
 as it is usually done in the framework of the $\overline{MS}$ scheme
 (this is also equivalent to a re-definition of $1/\omega$ poles).

 We can now calculate coefficients $S_n$ and $T_n$ of $1/\omega^n$
 poles in $S$ and $T$ including terms of the order of $g_s^2$:
 \be
 S_1&=&{N_{\! C}\over (4\pi)^2} \; {a^2+b^2\over
2m_W^2}\left(1+{g_s^2\over
 (4\pi)^2} \;
 s^{(1)}\right),
 \\[3mm]
 T_1&=&{N_{\! C}\over (4\pi)^2} \; {2abm_um_d+(a^2+b^2)(m_u^2+m_d^2)
 \over m_W^2m_H^2}\left(1+{g_s^2\over (4\pi)^2} \;
 t^{(1)}\right),
 \\[3mm]
 S_2&=&-{N_{\! C}\over (4\pi)^2} \;
 {a^2+b^2\over m_W^2} \; {4g_s^2\over (4\pi)^2},
 \\[3mm]
 T_2&=&-{N_{\! C}\over (4\pi)^2} \; {2abm_um_d+(a^2+b^2)(m_u^2+m_d^2)
 \over m_W^2m_H^2} \; {16g_s^2\over (4\pi)^2},
 \ee
 with $s^{(1)}=\frac{10}{3}$ and $t^{(1)}=\frac{8}{3}$.
 We have displayed $S_2$  and $T_2$ because they
 illustrate nice agreement of our calculation
 with equations following from the renormalization group analysis
 \cite{inami,gross}. Namely, both these quantities can be found
 in the lowest relevant order of perturbation theory from
 \footnote{Here and below, $\gamma_m^{(n)}$ and $\beta_n$ correspond
 to the coefficients of expansion (in $g_s$)
 of the anomalous dimension of mass, $\gamma_m(g_s)$, and the
 beta function, $\beta(g_s)$. In the normalization used (see, e.g.,
 ref.~\cite{inami} and references therein) we have:
 $\beta_0 = 11 - \frac{2}{3} N_F$,
 $\beta_1 = 102 - \frac{38}{3} N_F$,
 $\gamma_m^{(0)} = -8$,
 $\gamma_m^{(1)} = -\frac{404}{3} + \frac{40}{9} N_F$,
 where $N_F$ is the number of quark flavours.}
 \be
 S_2={g_s^2\over (4\pi)^2} \; \gzero \; S_1 + O(g_s^4), \hspace{1cm}
 T_2=2{g_s^2\over (4\pi)^2}\; \gzero \; T_1 + O(g_s^4),
 \ee
 and we reproduce our expressions by putting $\gzero=-8$.

 \section{Corrected decay width}
 \label{result}

 % Graph of the rate vs. mass of the Higgs
 Here we shall present the formula for the decay rate of the charged
Higgs
 boson
 including next-to-leading corrections. We define
  ${\cal L} = \ln \left( m_H^2/\Lambda_{\rm QCD}^2 \right)$ and
obtain:
 \be
 \lefteqn{
 \Gamma_H={N_{\! C}g^2m_H\over 32\pi m_W^2} \; \lll^{\gbe}}
 \nonumber\\[3mm]
 \hspace*{-8mm}&\times &\!\!\!\left\{ {\hat{a}^2+\hat{b}^2\over 2}
 \left[ 1+{\beta_1\gamma_m^{(0)}\over \beta_0^3} \; {\ln \lll\over
\lll}
 +{1\over \beta_0 \lll}
 \left(
2s^{(1)}-2\gzero+{1\over\beta_0}\left({\beta_1\over\beta_0}\gzero
 -\gamma_m^{(1)}\right)\right)\right]+\delta\right\} ,\nonumber\\ &&{
}
 \label{eq:mainres}
 \ee
 where the mass correction $\delta$ is
 \be
 \delta = {\lll^{\gbe}\over m_H^2}
 \left\{
 {2\gzero\over \beta_0 \lll}
 \left( 4\hat{a}\hat{b}\widehat{m}_u\widehat{m}_d
 +(\hat{a}^2+\hat{b}^2)(\widehat{m}_u^2+\widehat{m}_d^2) \right)
 \right. \hspace{4cm}
 \nonumber\\[3mm]
 -\left( 2\hat{a}\hat{b}\widehat{m}_u\widehat{m}_d
 +(\hat{a}^2+\hat{b}^2)(\widehat{m}_u^2+\widehat{m}_d^2)\right)
 \hspace{5cm}
 \nonumber\\[3mm]
 \times \left.
 \left[ 1+{2\beta_1\gamma_m^{(0)}\over \beta_0^3} \; {\ln \lll\over
\lll}
 +{2\over \beta_0 \lll}
 \left( t^{(1)}+{1\over\beta_0}\left({\beta_1\over\beta_0}\gzero
 -\gamma_m^{(1)}\right)\right) \right] \right\}.
 \label{eq:delta}
 \ee
 We have expressed the decay width of the charged Higgs boson
 in terms of the
 renormalization group invariant masses of the quarks $\widehat{m}_q
 \; (q = u,d)$, and
 the coefficients $\hat{a}$ and $\hat{b}$ defined by equations
 (\ref{eq:abI}) and  (\ref{eq:abII}) with $m_q$ replaced by
 $\widehat{m}_q$.\footnote{If we formally put
 $\widehat{m}_u = \widehat{m}_d = \hat{a} =\hat{b} \equiv
\widehat{m}_q$
 in eqs.~(\ref{eq:mainres}) and (\ref{eq:delta}), we get the correct
result
 for partial decay width of a neutral Higgs into $q\bar{q}$ pair
 (see eqs.~(3.21)--(3.22) in \cite{inami}).}
 Although eqs.~(\ref{eq:mainres}) and (\ref{eq:delta}) correspond to
 partial decay width $H^+ \rightarrow u\bar{d}$, it is clear that the
main
 contribution to the sum over generations will be given by the term
 with maximal quark masses allowed.

 Following the method of ref.~\cite{inami,georgipol} we use the
threshold
 condition stating that the running mass of the quark at the energy
scale of
 production of a pair quark-antiquark is equal to half this energy;
from
 this condition we obtain:
 \be
 \widehat{m}_q^{-2}  =  m_q^{-2}\left( \ln{4 m_q^2\over \Lambda_{\rm
QCD}^2}
 \right)^{\gbe}
 \hspace{85mm} \nonumber\\[3mm]
  \times\left\{1+{\beta_1\gamma_m^{(0)}\over\beta_0^3}
 {\ln\ln(4 m_q^2/\Lambda_{\rm QCD}^2) \over \ln(4m_q^2/\Lambda_{\rm
QCD}^2) }
 +{1\over \beta_0^2}
 \left(  {\beta_1\over\beta_0} \gamma_m^{(0)} -\gamma_m^{(1)}\right)
 {1 \over   \ln(4 m_q^2/\Lambda_{\rm QCD}^2) }    \right\} .
 \ee
 We now want to visualize the magnitude of the leading and
next-to-leading
 order corrections. The leading order correction can be obtained from
 equations (\ref{eq:mainres}) and (\ref{eq:delta}) by dropping all
the
 terms divided by  ${\cal L} = \ln \left(m_H^2/\Lambda_{\rm QCD}^2
\right)$.
 In the case of model II, our formulae
 in the leading logarithmic approximation reproduce the
 results of ref.~\cite{mend90}, for both reactions  $H^+\to c\bar s$
 and $H^+\to t\bar b$. To our knowledge, the corrections in model I
have
 not been analyzed so far even in the leading order. We present both
the
 leading and the next-to-leading logarithmic corrections
 to the rate of the decay $H^+\to t\bar b$
 in
 fig.~\ref{fig:widthvsmass}(a-c) for the value of $\tan\beta=2$, for
which
 this decay has similar rate in both models (we take $m_t$ = 150 GeV
and
 $m_b$ = 4.5 GeV).  It turns out that although the
 leading corrections decrease the expected width of the charged Higgs
boson,
 the next-to-leading terms can increase it by a large factor,
especially for
 the light mass of the Higgs. On the other hand, this effect may be
an
 artifact of the choice of the threshold condition. The sensitivity
of the
 result to this choice is shown in fig.~\ref{fig:widthvsmass}, where
we
 present the results using two different conditions:
 (a,b) $\overline{m}(4m^2)=m$ and (c) $\overline{m}(m^2)=m$
 (where $\overline{m}$ denotes
 the running mass of the quark). The dependence on the initial
condition
 has also been discussed in much detail in ref.~\cite{braa81}.

\begin{figure}[h]
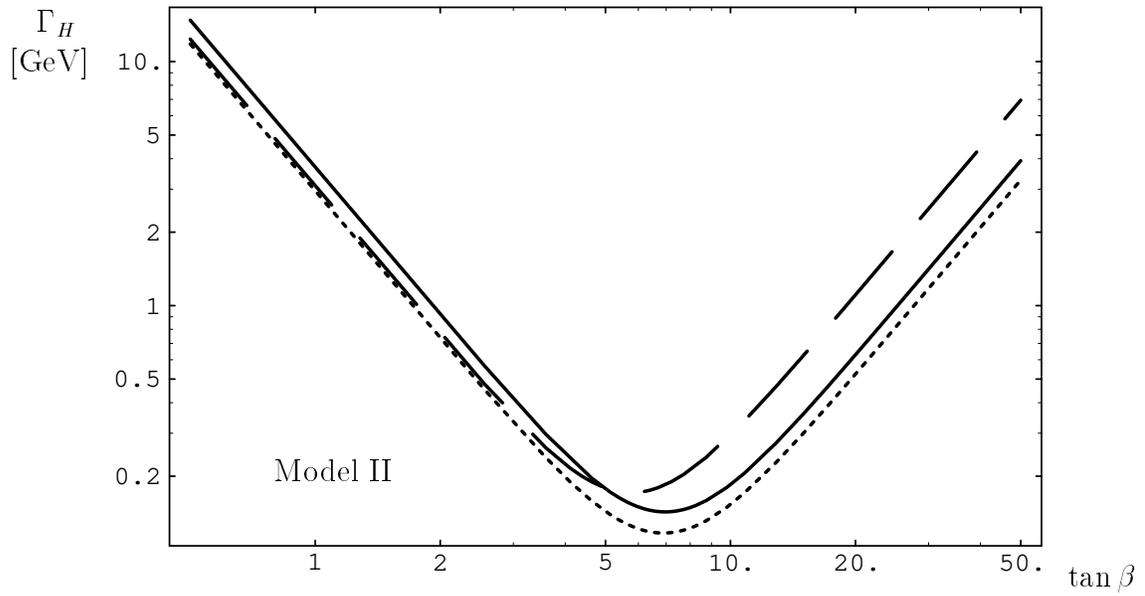

\caption{Rate of the charged Higgs decay as a function of
$\tan\beta$.
Dashed: Born rate, dotted:
leading, and solid: next-to-leading corrections. }
\label{fig:widthvstan}
\end{figure}

  Size of corrections also strongly depends on the value of
 $\tan\beta$, as can be seen in fig.~\ref{fig:widthvstan}.  For this
 plot we have used the usual condition $\overline{m}(4m^2)=m$, and we
see
 that
 for large $\tan\beta$, where the decay is dominated by the coupling
 proportional to the bottom quark mass, next-to-leading corrections
 slightly decrease the effect of the leading ones.  Therefore, the
effect
 of increasing the width mentioned above is due to the corrections to
the
 mass of the top quark; this may be a signal of insufficiency of the
 expansion in $m_q/m_H$ for lighter Higgs bosons.  We are going to
 address this issue in future (see also ref.~\cite{shirkov92}).

 \section{On the decay $t\rightarrow H^+ b$}
 \label{top}

    The same term in the Lagrangian which is responsible for the
decay of
 the charged Higgs boson into quarks will also enable a sufficiently
 heavy top quark to decay into a bottom quark and an $H^+$.  Both
 electroweak \cite{csli93,cza93} and QCD
\cite{liyuan90,liuyao92,sa92b}
 corrections to this decay have been studied at the one-loop level.
It
 has been found \cite{sa92b} that in the two Higgs doublet model
 predicted by supersymmetry the relative size of QCD corrections
becomes
 large for growing values of $\tan\beta$.  We would like to discuss
this
 effect here in order to demonstrate that this large correction can
be
 absorbed in the Born rate if one uses the running mass of the $b$
 quark\footnote{One of the authors (A.C.)
 is grateful to the referee of Physical Review D for
 suggesting this and to Sacha Davidson for
 helpful discussions on this topic.},
 just like in the case of the Standard Model Higgs
 boson \cite{bl80,braa81}.

\begin{figure}[h]
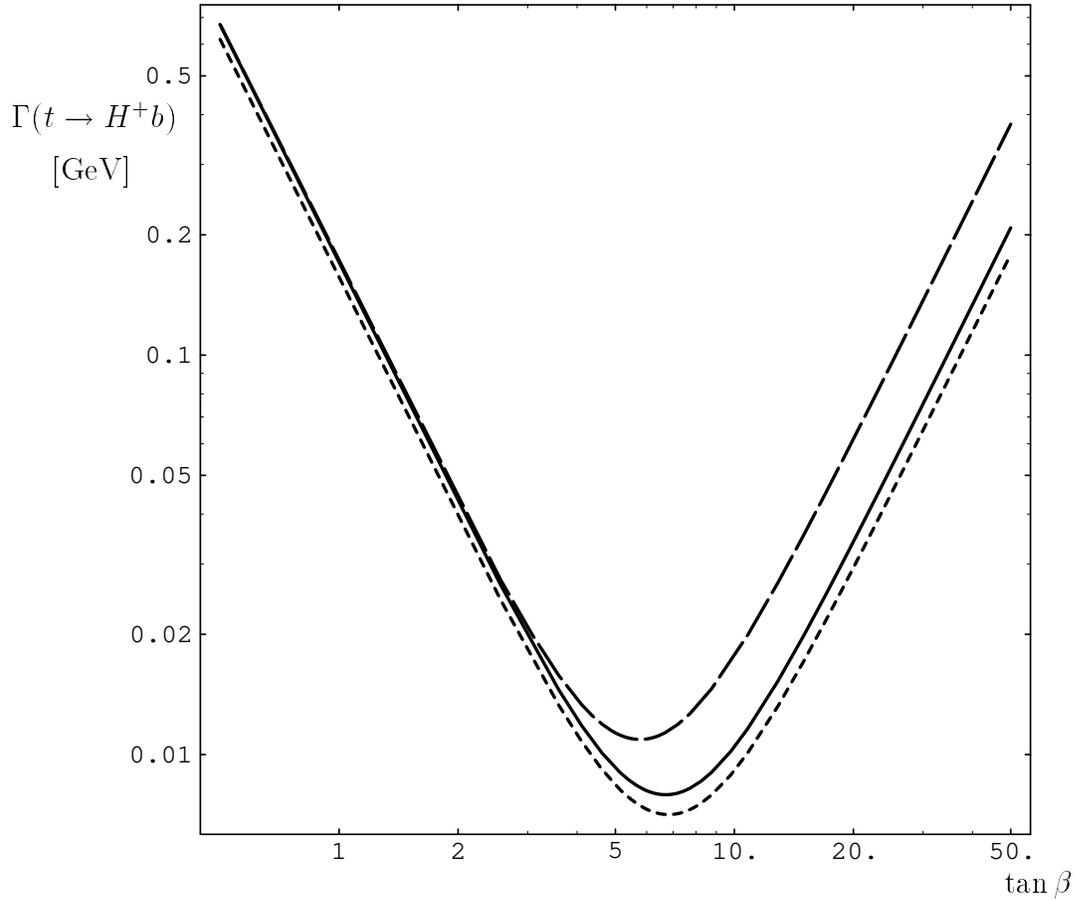

\caption{Rate of the decay $t \rightarrow H^+ b$ for $m_t=150$ GeV,
$m_H=80$
GeV, $m_b=4.5$ GeV and $\alpha_s=0.1$: Born rate (long dash),
rate including first order QCD corrections (short dash), and the
improved
Born rate (solid line).}
\label{fig:top}
\end{figure}

    For this purpose we compute the tree level decay rate in the
limit of
 very large $\tan\beta$ and mass of the top quark:
 \be
    \Gamma^{(0)} (t \rightarrow H^+ b)& =& \frac{G_Fm_t^3}{8 \sqrt2
\pi}
 | V_{tb}|^2 (1-\chi^2)\left[4+(1-\chi^2)\tan\beta\right]\epsilon^2 ,
 \ee
 where we have introduced the following notation for the ratios of
relevant
 masses:
  \be
 \epsilon={m_b\over m_t}, \qquad
 \chi={m_{H}\over m_t} \hspace{1cm} (m_{H} \equiv m_{H^+}).
 \ee
 The first order QCD corrections, calculated in ref.~\cite{sa92b},
can
 be expressed in the limit of large $\tan\beta$ by:
 \be
 \Gamma^{(1)}(t \rightarrow H^+ b)&=&{\alpha_s\over 6\pi}{G_F
 m^3_t \left|V_{tb}\right|^2 \over \sqrt{2}\pi}\left[ \left( 2G_+-G_0
 \right)\tan^2\beta +4 G_- \right]\epsilon^2,
 \ee
 and the explicit formulae for the coefficient functions $G_i$ can be
 found in ref.~\cite{sa92b}. Here we only need
 terms of the order of $\ln\epsilon$:
 \be
 G_+ &\rightarrow& {3\over 4}(1-\chi^2)^2\ln\epsilon,
 \nonumber\\[2mm]
 G_- &\rightarrow& 3(1-\chi^2)\ln\epsilon,         \nonumber\\[2mm]
 G_0 &\rightarrow& -{3\over 2}(1-\chi^2)^2\ln\epsilon.
 \ee
 Using these expressions we can calculate the asymptotic value of
first order
 corrections for large values of $\tan\beta$ and $m_t$:
 \be
 \Gamma^{(1)}={2\alpha_s\over \pi}\ln\left(\frac{m_b^2}{m_t^2}\right)
 \Gamma^{(0)}.
 \label{eq:asym}
 \ee
 We see that for $\alpha_s\approx 0.1$, $m_b=4.5$ GeV and $m_t=100$
GeV this
 correction is of the order of $-40\%$, in agreement
 with diagrams presented in \cite{sa92b}.
 The size of corrections
 becomes even
 larger as the mass of the top quark increases, and eventually the
one-loop
 corrected rate of decay becomes negative; such large corrections are
 a sign of a breakdown of the perturbation theory. However, it is
possible
 to avoid the large corrections by renormalizing the mass of the $b$
quark
 not
 on the mass-shell but at the energy scale
 characteristic to the process, which
 is the mass of the top quark. The running mass of the bottom quark
at this
 energy is:
 \be
 \overline{m}_b(m_t^2)=m_b\left(\ln{(4 m_b^2 / \Lambda_{\rm QCD}^2)}
 \over \ln{(m_t^2 /
 \Lambda_{\rm QCD}^2)}\right)^{12/(33-2N_F)},
 \label{eq:run}
 \ee
 where $N_F=6$ is the number of quark flavours, and we take
 $\Lambda_{\rm QCD}=150$ MeV (in the $\overline{\rm MS}$ scheme).
 We can expand the above expression in a series in the coupling
 constant $\alpha_s$, and we find that:
 \be
 \overline{m}_b(m_t^2)\approx m_b
 \left(1+{\alpha_s\over \pi}\ln\left( \frac{m_b^2}{m_t^2} \right)
\right).
 \ee
  It can now be seen from the formula (\ref{eq:asym}) that for large
 $\tan\beta$ and $m_t$ the one-loop corrected decay rate approaches
the
 Born rate expressed in terms of the running $b$ quark mass.
 In figure~\ref{fig:top}
 we show the dependence of decay rate of the top quark on
 $\tan\beta$. We note that for large values of $\tan\beta$ the QCD
corrected
 rate is not much different from the Born rate expressed in terms of
the
 running
 $b$ quark mass (\ref{eq:run}), and that we no longer face the
problem of
 unreasonably large corrections.

  On the other hand, for the
 small values of $\tan\beta$ the Born rate remains
 approximately unchanged when expressed in terms of the running $b$
mass. The
 reason for this is that the dominant coupling of the quarks to the
charged
 Higgs is $m_t\cot\beta$ in this region  and mass of the $b$ quark
does not
 play
 any important role. The same can be said about the analysis of the
top quark
 decay in the non-supersymmetric two Higgs doublet model, and it
explains
 why no
 large logarithmic corrections were found there for any values of
 $\tan\beta$~\cite{sa92b}.

 \section{Summary}    \label{summ}
    We have found leading and next-to-leading order corrections to
the
 decay width of the charged Higgs boson in the framework of two
models.
 In the case of the leading corrections in the model motivated by
 supersymmetry we confirmed the previously published formulae
 \cite{mend90}; the
 remaining results are new.  For a heavy Higgs boson or for large
values
 of $\tan\beta$ we found that the next-to-leading order corrections
 sizably decrease the effect of the leading order corrections, and
 increase the final result for the decay rate.  We have also examined
the
 process $t\to H^+b$, explaining the origin of large logarithmic
 corrections found in a previous paper \cite{sa92b}.

 \section*{Acknowledgment}
 A.C.'s research was  supported by  a  Dissertation
  Fellowship of the University of Alberta,  and by a grant to
Professor  A.N.
  Kamal  from  the  Natural  Sciences  and  Engineering  Research
Council of
 Canada.   A.D.'s research  was supported  by the Research Council
 of Norway.

 %\bibliographystyle{unsrt}
 %\bibliography{../dok/phd}

%\newpage

\section*{Figure captions}
\begin{figure}[h]
\caption{Rate of the charged Higgs decay (a) in Model II and (b,c) in
Model
I for two different threshold conditions (see text). Dashed: Born
rate,
dotted: leading, and solid: next-to-leading corrections.}
\label{fig:widthvsmass}
\end{figure}
\end{document}